\documentclass[useAMS,usenatbib]{mn2e}
\usepackage{epsfig}
\usepackage{amsmath}    
\usepackage{amssymb}
\title[The growth of supermassive black holes and LOFAR]{Probing the growth of
  supermassive black holes at ${\bf z>6}$ with LOFAR} \author[K.J. Rhook \& M.G.
Haehnelt]{Kirsty J.  Rhook$^{1}$\thanks{krhook@ast.cam.ac.uk} \& Martin G.
  Haehnelt$^{1,2}$\thanks{haehnelt@ast.cam.ac.uk}\\$^{1}$Institute of Astronomy,
  Madingley Road, Cambridge CB3 0HA\\$^{2}$Kavli Institute for
  Theoretical Physics,   Kohn Hall, UCSB, Santa Barbara CA 93106}
\begin{document}

\date{August 29th, 2006}

\pagerange{\pageref{firstpage}--\pageref{lastpage}} \pubyear{2005}
  
\maketitle

\label{firstpage}

\begin{abstract}
  HII regions surrounding supermassive black holes (BHs) in an
  otherwise still neutral intergalactic medium (IGM) are likely to be
  the most easily detectable sources by future 21cm experiments like
  LOFAR. We have made predictions for the size distribution of such
  HII regions for several physically motivated models for BH growth at
  high redshift and compared this to the expected LOFAR sensitivity to
  these sources.  The number of potentially detectable HII regions
  does not only depend on the ionisation state of the intergalactic
  medium and the decoupling of the spin temperature of the neutral
  hydrogen from the cosmic microwave background (CMB) temperature, but
  is also strongly sensitive to the rate of growth of BHs at high
  redshift. If the supermassive BHs at redshift $6$ were built up via
  continuous Eddington-limited accretion from low mass seed BHs at
  high redshift, then LOFAR is not expected to detect isolated QSO HII
  regions at redshifts much larger than 6, and only if the IGM is
  still significantly neutral.  If the high-redshift growth of BHs
  starts with massive seed BHs and is driven by short-lived accretion
  events following the merging of BH hosting galaxies then the
  detection of HII regions surrounding supermassive BHs may
  extend to redshifts as large as $8-9$ but is still very sensitive to
  the redshift to which the IGM remains significantly neutral. The
  most optimistic predictions are for a model where the supermassive
  BHs at $z>6$ have grown slowly.  HII regions around supermassive BHs
  may then be detected to significantly larger redshifts.

\end{abstract}

 
\section{Introduction}

Reionisation - the change of state of the intergalactic medium (IGM)
from a neutral state at recombination to the present day ionised state
- is an epoch which is currently under intense study. Analyses of the
spectra of high redshift QSOs indicate a smooth increase in the
optical depth of the IGM with redshift which accelerates at $z>6.1$
(Fan et al. 2002, Songaila 2004, Fan et al. 2006).  Quasars at $z>6.1$
show almost complete absorption of the flux blue-ward of the
Lyman$-\alpha$ resonance [the Gunn-Peterson (1965) effect].  Inferring
the ionisation state of the IGM at these high redshifts from QSO
absorption spectra is difficult and ambiguous (e.g. Oh \& Furlanetto
2005, Bolton \& Haehnelt 2006).  If the small amounts of flux 
transmitted just blue-ward of the
Lyman$-\alpha$ centre are attributed to the presence of a locally
ionised region embedded within a significantly neutral medium, then
lower limits on the neutral fraction of $x_{\rm HII}$ may be inferred
(Wyithe \& Loeb~2004a; Mesinger \& Haiman~2004). Malhotra \&
Rhoads~(2005), however, argue that the number density of
Lyman-$\alpha$ emitters can be used to place a lower limit on the
ionised fraction of gas at $z \sim 6$ at $20 - 50$~per cent. Fan et
al. (2006) also come to the conclusion that the IGM probed by the
highest redshift QSOs is predominantly ionised.

The optical depth to Thompson scattering inferred from the recent WMAP
measurements offers an independent constraint on the integrated
reionisation history. Interpreted as a sudden reionisation at a
redshift $z_{\rm reion}$ the year three WMAP results give $z_{\rm
reion} = 10^{+2.7}_{-2.3}$ (Spergel et al. 2006).

The emerging picture is consistent with the reionisation of cosmic
hydrogen being complete by $z \sim 6 - 13$ and therefore significant
contribution to reionisation by sources beyond current detection
limits.  Experiments to detect the redshifted 21cm
signature of high redshift neutral hydrogen will potentially provide
copious amounts of information on the epoch  of reionisation. The
European design for a low frequency interferometer, LOFAR, may be in
operation as early as 2007 and other low-frequency experiments (see
\textsection \ref{prospects}) are in various stages of planning and
development.  However, mapping the high redshift neutral gas
distribution in 21cm will be extremely challenging.

The redshifted 21cm signal due to ionised regions produced by bright,
high redshift QSOs surrounded by an otherwise still neutral
intergalactic medium may be our best bet for an detection of individual
objects due to their 21cm signature (see Tozzi et al. 2000 for a
detailed discussion).  Kohler et al.~(2005) used numerical simulations
to show that ionised regions around high redshift QSOs are indeed
likely to be the most prominent features in the high redshift
neutral gas distribution.  Wyithe, Barnes \& Loeb~(2005) use a
semi-analytic model for the high redshift QSO HII region distribution
and predict detectable HII regions at $z \sim 6 - 8$.  However, the
growth history of high redshift supermassive BHs is very uncertain. In
this paper we will investigate a range of models for high redshift BH
growth which reproduce the bright end of the luminosity function of
redshift $6$ QSOs and sample the range of plausible growth histories
at higher redshift.

We begin by reviewing the physics of detecting a 21cm signal due to
neutral hydrogen (\textsection \ref{21cm}). In section
\ref{QLF_models} we describe our models for the evolution of the QSO
luminosity function (QLF).  We discuss the expected size of HII
regions due to QSOs in \textsection \ref{HII_regions}.  Finally, in
\textsection \ref{prospects}, we compare our computed size
distributions to the expected sensitivity of
LOFAR\footnote{http://www.lofar.org/} to extended sources and discuss
observational prospects.

Throughout this paper we adopt a cosmological matter density $\Omega_m
= 0.27$, baryonic matter density $\Omega_b = 0.044$, cosmological
constant $\Omega_{\Lambda} = 0.73$, present day Hubble constant $H_o
\equiv  100h$~km~s$^{-1}$~Mpc$^{-1} = 71$km~s$^{-1}$~Mpc$^{-1}$, a mass
variance on scales of $8 h^{-1}$~Mpc $\sigma_8 = 0.84$ and a
scale invariant primordial power spectrum (slope $n=1$).

\section{Physics of 21cm emission/absorption}\label{21cm}

A ground state neutral hydrogen atom can emit or absorb at a
wavelength of 21cm corresponding to the energy of the
transition between the singlet and triplet hyperfine levels.  At the
low frequencies of this transition, the specific intensity of the
emission is proportional to the brightness temperature of
the source $T_b$ as described by the  the Raleigh-Jeans law, 

\begin{equation}
I(\nu) = \frac{2\nu^2}{c^2}k_B T_b.
\end{equation}

The brightness temperature of a typical patch of sky at redshift $z$
depends on the optical depth of the gas to 21cm radiation, the
temperature of the cosmic microwave background $T_{\rm CMB}$ and the
spin temperature of the gas,

\begin{equation}
T_b = T_{\rm CMB}e^{-\tau} + (1-e^{-\tau})T_s,
\end{equation}

\noindent where $\tau$ depends on the density and ionisation state of
the gas. This results in a differential brightness temperature, 

\begin{eqnarray}\label{deltaTb}\nonumber
\delta T_b &=& \frac{(T_b - T_{\rm CMB})}{T_{\rm CMB}(1+z)}\\ \nonumber
&\approx& 27~{\rm mK}~x_{\rm HI}\left(1-\frac{T_{\rm
CMB}}{T_s}\right) 
\left( \frac{\Omega_bh^2}{0.02}\right) \\
&&\left[ \left( \frac{1+z}{10}\right) \left( \frac{0.27}{\Omega_m}\right)\right]^{1/2}.
\end{eqnarray}

Significantly neutral hydrogen gas for which $T_s$ differs from
$T_{\rm CMB}$ will emit or absorb at 21cm in contrast to the CMB.  The
value of $T_s$ is determined by the kinetic temperature of the IGM,
$T_{\rm IGM}$, and the strength with which $T_s$ and $T_{\rm IGM}$ are
coupled via collisional ionisation and/or Lyman$-\alpha$ pumping.
$T_{\rm IGM}$ is expected to decouple from $T_{\rm CMB}$ as early as
$z \sim 150 - 200$ as the timescale for Compton scattering off
residual free electrons becomes longer than the expansion timescale
(e.g. Scott \& Rees~1990). In the absence of a heat source, the IGM
cools adiabatically below $T_{\rm CMB}$ and therefore the very first
(Population-III) sources are expected to be born in a cold IGM.

The redshift range over which HII regions will appear in absorption is
uncertain; if $T_s$ couples to $T_{\rm IGM}$ prior to significant long
range X-ray heating there may be a significant absorption epoch.
Furlanetto (2006) argues  that if the (thermal) radiation from Pop-II
stars dominate reionisation then this will be the case, but if Pop-III
stars were dominant in reionising the universe then the IGM could  be
heated above $T_{\rm CMB}$ before there are sufficient Lyman$-\alpha$
photons to couple $T_s$ to $T_{\rm IGM}$ (see also Chen \&
Miralda-Escud{\' e}~2003 and Kuhlen, Madau \& Montgomery 2006). 
For simplicity we assume here that the temperature of
the IGM lies well above the temperature of the CMB.  Regions of
ionised gas then appear as holes in the 21cm emission due to
the IGM (see also Kohler et al. (2005) for a more  detailed discussion). 

With the rise of the galaxy and QSO populations the ionised
fraction of gas increases and the 21cm signal emission due to the IGM
becomes weaker. The signal will disappear by $z < 6$ when we know the
neutral hydrogen is predominantly ionised.  There will be a finite
period of time when the IGM is heated above the temperature of the CMB
but most of the gas is still neutral.  This is the (uncertain)
redshift range over which instruments like LOFAR can probe.  In
addition, the performance of LOFAR will depend on our ability to model
and remove foreground contamination due to the ionosphere, our own
galaxy, and both extended and point-like extragalactic sources.

\section{Models for the luminosity function of high redshift QSOs}\label{QLF_models}

\subsection{The growth of supermassive black holes at high redshift}

We are interested here in the HII regions around supermassive BHs at
high redshift. Accretion onto a BH is an efficient mechanism for
producing ionising photons and therefore a supermassive BH that has
been built up by gas accretion should generate a very large ionised
bubble. If the ionising flux generated by the supermassive BHs at high
redshift exceeds that produced by other sources in the same region
then the sites of supermassive BHs will show up prominently in 21cm
maps as large ionised regions in an otherwise still neutral IGM. In
order to estimate the space density and size distribution of these HII
regions we require a model for the growth of the supermassive black
holes and the production of ionising photons. Our best anchor point
for such modelling is the observed luminosity function of $z=6$ QSOs.
We therefore consider here models which reproduce this luminosity
function but differ in the timescale over which the black holes
powering the $z=6$ QSOs grow.

\subsection{Passive evolution model}\label{lum_model}

The e-folding time for BH growth at the Eddington limit is $\sim 0.44
(\epsilon_{\rm acc}/(1-\epsilon_{\rm acc})) $~Gyr, where
$\epsilon_{\rm acc}$ is the fraction of the mass energy released as
radiation.  Estimates for the mass of seed BHs produced by
Population-III remnants are $\sim 10 - 10^{3}$~M$_{\odot}$. Such seed
masses would require $\sim 15 - 20$ e-foldings to grow to $3\times
10^{9}$~M$_{\sun}$, the estimated mass of the black holes powering the
observed $z=6$ QSOs.  The growth of a $3\times 10^9$~M$_{\odot}$ BH by
a redshift $\sim 6$ from a Population-III remnant therefore requires
more or less continuous accretion if the accretion rate is limited to
the Eddington accretion rate. It is a matter of intense debate whether
accretion above the Eddington rate contributes significantly to the
growth of supermassive black holes (see Begelman, Volonteri \& Rees
2006 for a recent discussion).

We first construct a  simple passive evolution model for the growth of 
supermassive black holes and the evolution of the QSO luminosity
function at $z > 6$ in which a fixed comoving density of black holes 
accrete at their Eddington limit with a fixed duty cycle $f_{\rm duty}$. 

The bolometric luminosity of a QSO powered by a BH of mass $M_{\rm
  bh}$ radiating at the Eddington limit is  given by 
\begin{eqnarray}\nonumber
L_{\rm bol} &=& \frac{4 \pi G c m_{\rm p} M_{\rm bh}}{\sigma_e},\\ 
&=& 3.28\times10^4 {\rm L}_{\odot} \frac{M_{\rm bh}}{M_{\odot}} 
\equiv \mathcal{L} {\rm L}_{\odot} \frac{M_{\rm bh}}{{\rm M}_{\odot}},
\end{eqnarray}

\noindent where 
$\sigma_e$ is the Thompson scattering cross section. The B-band
luminosity is assumed to be a fraction $f_B = 1/10$ of $L_{\rm bol}$.
Assuming that a fraction $\epsilon_{\rm acc} = 0.1$ of the mass energy
of the accreted mass is released as radiation, the ensemble average
luminosity evolution with respect to the redshift $6$ population is
then given by,

\begin{eqnarray}
L(z) &=& L(z=6) e^{-(t_{z=6}-t_z)f_{\rm duty}/\kappa},\\ 
\kappa &=& \frac{c \sigma_e}{4\pi G m_{\rm p}}\frac{\epsilon_{\rm acc}}{1-\epsilon_{\rm acc}},\\ \nonumber
&\simeq& 5 \times 10^8~{\rm yrs}\frac{\epsilon_{\rm acc}}{1-\epsilon_{\rm acc}},
\end{eqnarray}

\noindent where $t_z \equiv
t(z)$ is the cosmic time at redshift $z$.  

We adopt the QLF of \textsection \ref{merger_model} as our template
QLF at $z=6$.  In this model all luminosities grow exponentially and
the shape of the luminosity function is independent of redshift.  We
consider two values of the total duty cycle $f_{\rm duty} = [0.1,1.0]$.
To account for a possible contribution to the growth in an optically
obscured accretion mode we split the duty cycle of accreting black
holes into phases of luminous growth ($f_{\rm lum}$) and growth in an
obscured phase ($f_{\rm obs}$).  We assume the QSO spends equal
amounts of time in each phase so that $f_{\rm lum} = f_{\rm obs} =
[0.05, 0.5]$. The observed lifetime of a QSO in this model is
redshift dependent, and depends on the duty fraction of the luminous
population,

\begin{eqnarray}\nonumber
t_q &=& f_{\rm lum}t_H(z) \\
&\simeq& 3.5 \times 10^8~{\rm yrs}~\left(\frac{f_{\rm
lum}}{0.1}\right) \left(\frac{1+z}{7}\right)^{-3/2}.\\
&&~{\rm for}~z>>1. \nonumber
\end{eqnarray}

\noindent Here $t_H(z) = \frac{2}{3H(z)}$ is the Hubble time at a
redshift $z$ and $H(z) = H_o [(1+z)^3
\Omega_m+\Omega_{\Lambda}]^{1/2}$ is the Hubble parameter.

\begin{figure*}\label{plot1}
  \vspace*{75mm} \includegraphics{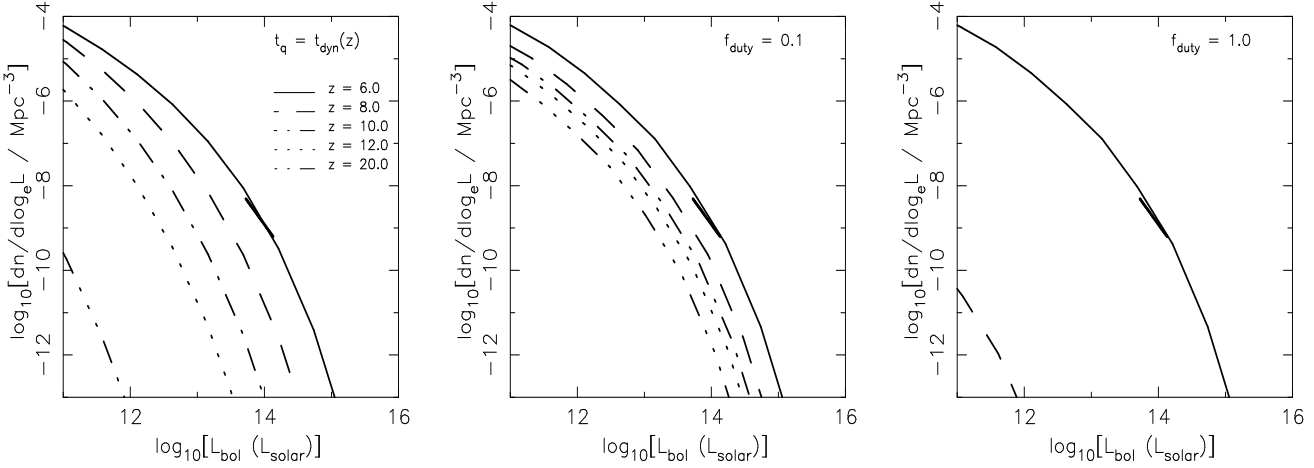}
\caption{\label{fig1} The thin curves show the QLF as a function of 
  bolometric QSO luminosity at $z=6,8,10,12,20$ for the merger rate
  model with $\epsilon_o = 10^{-4.8}$ (\emph{left} panel), and the
  passive evolution models with obscured duty fractions equal to the
  duty fraction of the luminous population and total duty fraction
  $f_{\rm duty} = 0.1$ (\emph{middle} panel) and $f_{\rm duty} = 1.0$
  (\emph{right} panel). The thick line shows the Fan et al.~(2004) fit
  to the observed $z \sim 6$ luminosity function at a rest frame
  wavelength of $1450 \AA$, where we have assumed that 10 per cent of
  the energy is radiated in the B-band and converted to a B-band
  magnitude using $M_B = M_{1450} - 0.48 + 5log_{10}\left(
    \frac{h}{0.65} \right)$. Note that the dashed line in the bottom
  left corner of panel 3 is the $z=8$ QLF.}
\end{figure*}

\subsection{A merger driven model}\label{merger_model}

\subsubsection{Galaxy mergers as triggers for QSO activity}

The activity powering optically bright QSOs is thought to be triggered
by galaxy mergers, which drive cold gas onto a central black hole.
Various models for the growth of black holes in hierarchically merging
galaxies have been built (e.g. Kauffmann \& Haehnelt~2000, Volonteri
et al.~2003) and have been successful in reproducing the observed
QSO luminosity function at $z<5$ . These models generally assign a
QSO light curve to each merging event, guided by observationally
and/or physically motivated estimates for the central black hole mass
and associated accretion rate (see Hopkins et al. 2005 for light
curves motivated by numerical simulations of a galaxy merger).  Below
we construct a simple version of such a model for the QLF which uses
the (extended) Press-Schechter formalism for dark matter halos to
estimate the rate of galaxy mergers (see Erickcek et al. (2006) for a
recent critical assessment of the accuracy of the extended
Press-Schechter formalism).  The model is similar to those described
by Haehnelt, Natarajan \& Rees (1998) and Wyithe \& Loeb~(2003).

\subsubsection{Estimating merger rates of black holes}

Our model makes a number of simplifying assumptions. 
We assume that the merger rate of black holes is equal to the 
merger rate of the dark matter halos of the galaxies hosting the
black holes as estimated using the (extended) Press-Schechter
formalism.   We assume a relation of the mass of the BH, $M_{\rm bh}$,
and the mass of  the dark matter halo forming in a merger $M$ as 
motivated by the observed correlation between $M_{\rm bh}$ and the 
velocity dispersion of the host galaxies stellar bulge 
/dark matter halo,

\begin{equation}\label{Msigma}
M_{\rm bh} \propto \sigma^{\alpha},
\end{equation}

\noindent in nearby galaxies (Gebhardt et al. 2000,
Ferrarese \& Merritt 2000,  Ferrarese~2002). 
This relation appears to hold independently of redshift for
QSOs out to $z \sim 3$ (Shields et al. 2003, 2006). Empirical
estimates of $\alpha$ typically fall in the range $4-5$. We have
chosen $\alpha = 5$ which is  consistent with a simple
self-regulated growth scenario in which the black hole grows until it
radiates enough energy to unbind the gas that is feeding it (e.g.
Silk \& Rees 1998, Haehnelt Natarajan \& Rees 1998, Wyithe \& Loeb
2003). We further assume that the velocity dispersion of the
stellar bulge is proportional to the circular velocity of the dark
matter halo at the virial radius, $\sigma = v_{\rm vir}/\sqrt{2}$,
with (Barkana \& Loeb 2001),

\begin{eqnarray}\label{vvir}
\nonumber 
v_{\rm vir} &=& 420 \left( \frac{M}{10^{12} h^{-1}{\rm
M}_\odot} \right)^{1/3} \left[ \frac{\Omega_m}{\Omega_m^z}
\frac{\Delta_c}{18\pi^2}\right]^{1/6} \\
&\times& \left( \frac{1+z}{7} \right)^{1/2}{\rm kms}^{-1},
\end{eqnarray}

\noindent where
$\Omega_m^z=\frac{\Omega_m(1+z)^3}{\Omega_m(1+z)^3+
  \Omega_\Lambda+\Omega_k(1+z)^2}$, $d \equiv \Omega_m^z-1$ and
$\Delta_c=18\pi^2+82d-39d^2$ is the overdensity of a virialised halo
at redshift $z$. Equations (\ref{Msigma}) and (\ref{vvir}) lead to a
dependence of black hole mass on host halo mass $M$ and redshift
of the form,

\begin{eqnarray}\label{Mbh-Mhalo}\nonumber
M_{\rm bh} &=& \epsilon(M,z) M\\
&=& \epsilon_o h^{\alpha/3} \left[ \frac{\Omega_m \Delta_c}{\Omega_m^z 18 \pi^2} \right]^{\alpha/6}\\\nonumber
&&\times (1+z)^{\alpha/2} \left(\frac{M}{10^{12} M_{\sun}}\right)^{\alpha/3 - 1}M.\\\nonumber
\end{eqnarray}

\noindent Note that $\epsilon_o$ sets the ratio of $M_{\rm bh}/M$ for a given
redshift.

Assuming that all dark matter halos host a central BH with mass given
by equation~(\ref{Mbh-Mhalo}), the merging rate of halos of mass
$(M-\Delta M)$ and $\Delta M$ is equal to the merger rate of halos
hosting BHs of mass $(M_{\rm bh} - \Delta M_{\rm bh})$ with those
hosting BHs of mass $\Delta M_{\rm bh}$ for $\Delta M_{\rm bh} =
\epsilon(\Delta M, z(t)) \Delta M$ and $M_{\rm bh} = \epsilon(M, z(t))
M$. $N_{\rm halo}(M, \Delta M, t)$ is given by the product of the
probability per unit time that a halo mass $\Delta M$ will merge with
another halo to form a halo with mass $M$ and the space density of
halos with the appropriate mass difference,

\begin{eqnarray}\label{bh_mergers}
N_{\rm halo}(M,\Delta M,t) &=& \left.\frac{d^2P}{d \Delta M dt} \right|_{M - \Delta M}\frac{dn}{d(M- \Delta M)}\\ \nonumber
 &=& N_{\rm bh}(M_{\rm bh},\Delta M_{\rm bh},t).
\end{eqnarray}

\noindent The mass density is derived from the Press-Schechter
(1974) halo mass function with the modification described by Sheth \&
Tormen~(1999) and the halo merger probability is taken from the
extended Press-Schechter formalism (Lacey \& Cole~1993).

\subsubsection{From merger rate to luminosity function}

\noindent We assume that following a merger, each black hole accretes and
radiates at its Eddington limit for a period equal to the dynamical
time of gas in a disk of radius $0.035$ times the virial radius as in 
Wyithe \& Loeb~2003, 

\begin{eqnarray}\label{tdyn}\nonumber
 t_q &=& t_{\rm dyn}(z)\\
 &=& 10^7~{\rm
  yrs} \left[ \frac{\Omega_m}{\Omega_m^z}
  \frac{\Delta_c}{18\pi^2}\right]^{-1/2} \left( \frac{1+z}{3}\right)^{-3/2},
\end{eqnarray}

\noindent The light curve is assumed to be a simple
top-hat function $L(M_{\rm bh},t)/{\rm L}_{\odot} = \mathcal{L}M_{\rm
  bh}/{\rm M}_{\odot} \Theta [t_q-(t-t_z)]$ (with $\Theta$ the usual Heaviside step function).

The QLF can then be estimated by integrating over the merger rate of
all BH pairs as follows,

\begin{eqnarray}
\Phi(L,z) &=& \int_{0}^{\infty} d M_{\rm bh}\int_{0.25\epsilon M}^{0.5\epsilon M} d\Delta M_{\rm bh} \int_{z}^{\infty} dz^{\prime} \frac{dt^{\prime}}{dz^{\prime}}\\ \nonumber 
&& \times N_{\rm bh}(M_{\rm bh}, \Delta M_{\rm bh}, t^{\prime}) \delta[L - L(M_{\rm bh},t)].
 \end{eqnarray} 
 
 If we perform the integral over $M_{\rm bh}$ using the $\delta$
 function, use equation~(\ref{Mbh-Mhalo}) to make the change of
 variable from BH masses to halo masses and assume that the merger
 rate changes little over the QSO lifetime [i.e.  $t_q << t_H(z)$],
 we get (see Wyithe \& Loeb 2003),

\begin{eqnarray}
\Phi(L,z) &=& \int_{0.25M}^{0.5M} d\Delta M N(M,\Delta M, t(z))t_q \frac{3}{\alpha \epsilon} \\ \nonumber
&& \frac{1}{\mathcal{L}~{\rm L}_{\odot}~{\rm M}_{\odot}^{-1}},
\end{eqnarray}

\noindent where $L/{\rm L}_{\sun} = \mathcal{L} \epsilon(M,z)M / {\rm M}_{\sun}$. 
Note that the integral over $\Delta M$ is truncated at the lower end
to ensure that we are only counting major mergers and at the upper end
to count each potential halo merger pair only once.

In principle, $\alpha$, $\epsilon_o$ and $t_q$ may be tuned to fit our
model QLF to the observed QLF at $z \sim 6$.  We have fixed $\alpha =
5$ and $t_q = t_{\rm dyn}(z)$ and varied $\epsilon_o$ to fit to the
observed luminosity function. For $\epsilon_o = 10^{-4.9}$ we obtain
an acceptable fit to the observed luminosity function (see Figure~1).
This normalisation is consistent with the corresponding empirical
normalisation of equation (\ref{Msigma}) for galaxies in the local
universe ($\epsilon_o \approx 10^{-4.9}$, see Ferrarese~2002).

Note that with the assumption $t_q = t_{\rm dyn}$ the QSO lifetime
at $z = 6$ is $\approx 1.3 \times 10^7$~yrs, which is consistent with
other estimates of the duration of the optically bright phase of QSOs
(Haehnelt, Natarajan \& Rees 1998, Haiman \& Hui 2001, Martini \&
Weinberg~2001, Hopkins et al.~2005).  Note that the shorter lifetime
in the merger driven model compared to the passive evolution model
results in smaller HII regions for the assumed values of $f_{\rm
  duty}$.

\subsection{Comparison with observations}\label{model_comparison}

\subsubsection{Quasar luminosity function}

Figure 1 shows the evolution of our model QSO luminosity function at
$z>6$ (the passive evolution model with two values of the duty cycle
and the merger driven model).  By construction all models agree with
the observed space density of bright QSOs at $z \sim 6$ as determined by
Fan et al. (2004). The redshift evolution differs dramatically for the
three models.  Fan et al.~(2001) find that the observed space density
of bright QSOs decreases with redshift like $e^{-1.15z}$ for
$3.6<z<6$. We compare this to the redshift evolution in our model QLFs
by summing over the bright end of each QLF. We sum from a minimum
luminosity $L_{\rm bol} = 5.3 \times 10^{13}$~L$_\odot$ which
corresponds to a rest frame $1450\AA$ absolute magnitude of $-26.8$
[for a conversion to B-band magnitude $M_B = M_{\rm 1450} - 0.48 +
5log_{10}\frac{h}{0.65}$ (Fan et al. 2001)].  The results are
displayed in Figure~2. For the merger driven model the density of
QSOs declines with a similar slope to the decline in the observed
space density at lower redshift.  In the passive evolution model the
rate of decline depends sensitively on the assumed duty fraction.  For
the lower value of the assumed duty cycle ($f_{\rm duty} = 0.1$) the
black holes grow relatively little at high redshift and the drop in
the space density of bright QSOs is significantly slower than in the
merger driven model.  For the high value of the duty cycle ($f_{\rm
  duty} = 1$) the QSOs are growing rapidly and quickly decline in
luminosity at higher redshift. As we will discuss in more detail below
the three models span a wide range of possible growth histories of
supermassive black holes at high redshift.  We should also note that
all three models overpredict the space density of bright QSOs at lower
redshifts. As discussed in detail by Bromley, Somerville \& Fabian
(2004) the rapid growth history of supermassive black holes at high
redshift is not easily reconciled with the much slower observed growth
at low redshift which is normally modelled with rather severe feedback
effects due to energy input by stars and/or the black hole. As we are
only interested in the growth of supermassive black holes before
reionisation we make no attempt to model this here.

\begin{figure}\label{plot2}
  \vspace{85mm} \includegraphics{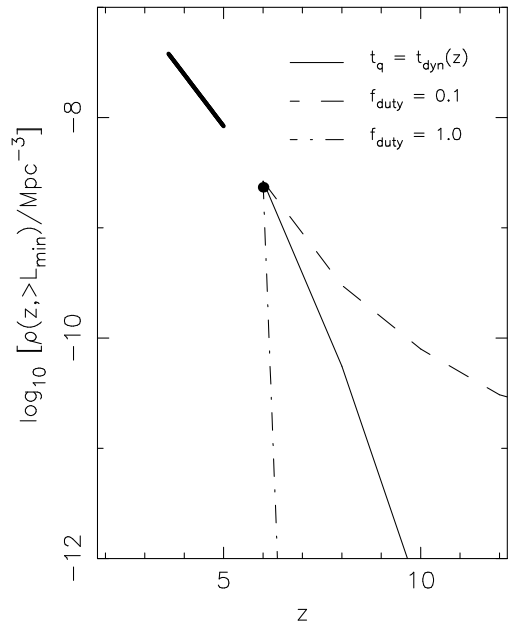}
  \caption{\label{fig2} Redshift evolution of QSO space density 
    for each of the three models plotted in Figure 1; the merger rate
    model with lifetime of $t_q = t_{\rm dyn}(z)$ from equation
    (\ref{tdyn}) (\emph{thin solid} line), the passive evolution model
    with fixed duty cycle $f_{\rm duty} = 0.1$ (\emph{dashed} line)
    and $f_{\rm duty} = 1.0$ (\emph{dot-dashed} line). The \emph{solid
      thick} line shows the evolution found for SDSS QSOs in the
    redshift range $3.6<z<5$ (see Fan et al. 2004) and the \emph{dot}
    is the estimated space density of QSOs at $z = 6$ (Fan et
    al.~2004). We have taken $L_{\rm min} \approx
    5 \times 10^{13}$~L$_{\odot}$ which is the bolometric luminosity
    corresponding $M_{1450} = -26.8$ assuming that $M_B = M_{1450} -
    0.48 + 5log_{10}(\frac{h}{0.65})$ and ten per cent of the
    bolometric luminosity is radiated in the B-band.}
\end{figure}

\subsubsection{Black hole mass densities and accretion rates}

The  accretion rate of mass onto black holes during the optically
bright phase of accretion can be extracted
from the optical QLF as follows, 

\begin{equation}
\frac{d\rho_{\rm acc, lum}}{dz}dz = \int_{z+dz}^{z} dz^{\prime}
\frac{dt}{dz^{\prime}}\int dL\frac{dn}{dL}(z^{\prime})L(z^{\prime})
\frac{1-\epsilon_{\rm acc}}{\epsilon_{\rm acc}c^2}. 
\end{equation}

An approximation to the total mass density in BHs can be extracted
from the model luminosity function by dividing the luminosity function
by the fraction of luminous/active halos at a given time $f_{\rm
  lum}(L,z)$,

\begin{equation}
\rho_{\rm bh}(z) = \frac{1}{\mathcal{L}~{\rm L}_{\odot}~{\rm M}_{\odot}^{-1}}\int dL \frac{dn}{dL}\frac{L}{f_{\rm lum}(L,z)}.
\end{equation}

For the passive evolution model $f_{\rm lum}$ is assumed to be
constant.  Note, however, that this will somewhat overestimate the BH
mass density. For the merger rate model $f_{\rm lum}$ is estimated to
be the ratio of the QSO lifetime to the average survival time of a
dark matter halo as a function of halo mass and redshift [$t_s(M,z)$]
calculated from the probability distribution in equation~(2.21) of
Lacey \& Cole~(1993).  At low redshift $t_s$ is approximately the
Hubble time, but it decreases to 30 per cent of the Hubble time by $z
\sim 12$ for the relevant halo masses. Figure 3 shows the evolution of
the black hole mass density and the rate of mass accretion during
optically bright phases for our three models.  As expected the
evolution differs strongly between the three models. We also show the
observed black hole mass density at $z=0$ which is about $30-300$
times larger than our estimates at $z=6$.  The difference in the
integrated observed accretion rate and the mass density in BHs is due
to the assumed periods of obscured QSO accretion.  By construction the
integrated luminous accretion rate will fall short of the mass density
in black holes in the passive evolution models by a factor $f_{\rm
  duty}/f_{\rm lum}=2$ at $z = 6$.  In our merger driven model this
difference is much larger.  Approximately ten per cent of the inferred
BH mass density is accumulated due to luminous accretion. In this
model approximately 90 per cent of the growth would have to occur in
an obscured phase. Such a growth scenario is consistent with the
findings of Hopkins et al. (2005) based on hydrodynamical simulations
of QSO activity fuelled by merging disk galaxies.

\begin{figure}\label{plot3}
  \vspace{85mm} \includegraphics{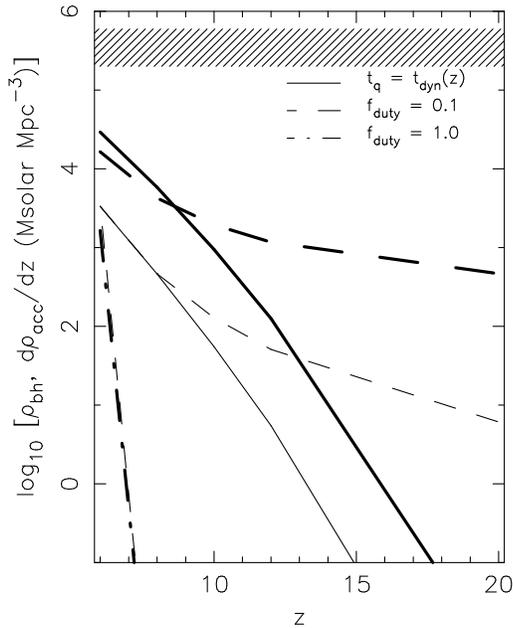}
  \caption{\label{fig3} Redshift evolution of the mass density in black holes 
    (\emph{bold} lines) and observed accretion rate of matter onto
    black holes (\emph{thin} lines) as described in
    \textsection\ref{model_comparison} for the three models described
    in \textsection\ref{QLF_models}. The \emph{hatched} regions shows
    the range of estimates for the mass density in BHs at $z=0$ (Yu \& Tremaine~2002).}
\end{figure}

\section{HII regions around high redshift supermassive black holes}\label{HII_regions}

Assuming that the QSO is embedded in a homogeneous gas distribution
with neutral hydrogen number density $\overline{n}_H = x_{\rm HI}
\overline{n}_{H,o}(1+z)^3$, the increase of the physical volume of ionised
hydrogen  $V_{\rm HII}$ generated by an ionising flux $\dot{N}_{\rm ion}(t)$ is given by
(e.g. Madau, Haardt \& Rees~1999, hereafter MHR99),

\begin{equation}\label{ode}
\frac{dV_{\rm HII}}{dt} - 3H(t)V_{\rm HII} = \frac{\dot{N}_{\rm
ion}(t)}{\overline{n}_H(t)} - \frac{V_{\rm HII}}{\overline{t}_{\rm rec}(t)},
\end{equation}

\noindent We assume the cosmic abundance of hydrogen 
with a neutral fraction $x_{\rm HI}$ and that the temperature of the
ionised IGM ($T_{\rm IGM}$) is $10^4$~K. The average hydrogen
recombination timescale $\overline{t}_{\rm rec}$ for such a gas is
(see Abel et al.~1997),

\begin{equation}
\overline{t}_{\rm
  rec} = 3 \times 10^8{\rm yrs} \left( \frac{T_{\rm IGM}}{10^4~{\rm
      K}} \right)^{0.7} \left[ \frac{1+z}{7} \right]^{-3}\left( \frac{C_{\rm eff}}{3}\right)^{-1},
\end{equation}

\noindent where we have assumed an effective clumping factor $C_{\rm
  eff} = 3$.  The effective clumping factor relevant for recombinations
 is expected to be low since the densest  self-shielded regions of neutral gas 
 have a small covering factor and should  not impede the progression of
 the ionisation front (see Miralda-Escud{\'e}, Haehnelt \& Rees~2000
  for a detailed discussion).

Equation (\ref{ode}) is integrated numerically to solve for the
physical radius of the HII regions for QSOs in each
of our QLF models,

\begin{equation}\label{size}
R_{\rm HII} = \left( \frac{3}{4\pi} \int_{t(z) - t_q}^{t(z)}
\frac{dV_{\rm HII}}{dt} \right)^{1/3}. 
\end{equation}

\noindent Equation (\ref{ode}) is approximately equivalent to
assuming that the volume of the ionised region is equal to the
ratio of the number of emitted ionising photons and the density of
neutral hydrogen,

\begin{equation}
V_{\rm II} \approx \frac{<\dot{N}_{\rm ion}>t_q}{\overline{n}_{H}}.
\end{equation}

\noindent For a QSO with constant luminosity, this approximation
breaks down as $t_q \rightarrow \overline{t}_{\rm rec}$ and the HII
region reaches its Str$\ddot{\rm o}$mgren radius. However for a QSO
with luminosity growing exponentially with the mass of the central BH,
most of the HII region growth occurs between $t(z) - \kappa$ and
$t(z)$.  Since $\kappa \sim \overline{t}_{\rm rec}$, the result
holds for the QSOs with $t_q > \overline{t}_{\rm rec}$ in our
models. Note that since little ionising flux is produced early in the
lifetime of a QSO powered by a BH continually accreting at the
Eddington limit,  the size of the HII regions at $z = 6$ are 
approximately equal for each of the passive evolution models 
even though the lifetimes differ by a factor of 10.

For each of our models the radial extent of the ionised region
generated by a QSO depends inversely on the surrounding neutral
hydrogen fraction $x_{\rm HI}$ and increases with the QSO
luminosity as $R_{\rm HII} \propto L_{\rm bol}^{1/3} x_{\rm HI}^{-1/3}$.

We calculate $\dot{N}_{\rm ion}$ at all redshifts by assuming the
optical/UV spectral energy distribution [$L(\nu)$] as given by MHR99
with a high energy cut-off of $200$~keV.  The spectral
energy distribution is normalised by assuming $\nu_B L_B = f_B L_{\rm
  bol}$ (with $f_B = 0.1$ as in \textsection \ref{QLF_models}). This
gives an ionising flux,

\begin{equation}
  \dot{N}_{\rm ion} = \int_{\nu_{\rm ion}}^{\infty}\frac{L(\nu)}{h\nu} d \nu \approx 1.5 \times 10^{57}~{\rm s}^{-1}\left(\frac{L_{\rm bol}}{10^{14}L_{\odot}}\right),
\end{equation}

\noindent where $\nu_{\rm ion}$ is the minimum photoionising
frequency.  The corresponding radius of the HII region, for a QSO with constant luminosity is,

\begin{eqnarray}\label{lum-rad}
R_{\rm HII} &=& 3.6 ~{\rm Mpc} \left(x_{\rm HI}\right)^{-1/3}
\left(\frac{L_{\rm bol}}{10^{14}~L_{\odot}}\right)^{1/3} \\ \nonumber
 &&\times \left[\frac{t_q}{1 \times 10^7 {\rm yrs}}\right]^{1/3}
\left(\frac{1+z}{7}\right)^{-1}.
\end{eqnarray}

\section{Prospects for detecting high redshift HII regions}\label{prospects}

\subsection{Future high-redshift 21cm experiments} 

The next generation of radio telescopes are aiming to explore the
low-frequency universe, with detection of an epoch of reionisation
signal a major scientific goal.  Various low-frequency interferometers
are being designed, tested and built with some already taking data.
The largest radio telescope on the horizon is the Square Kilometre
Array (SKA\footnote{http://skatelescope.org/}). The SKA is still in
the design phase and not expected to be in operation for more than a
decade.  However there are several smaller-scale telescopes including
LOFAR (to be built in the Netherlands), the Mileura Wide-Field Array
(MWA \footnote{http://haystack.mit.edu/arrays/MWA}, sited in Western
Australia) and the Primeval Structure Formation telescope (PAST
\footnote{http://web.phys.cmu.edu/past/}, currently being built in
China) that are expected to be in operation on much shorter
time-scales. These arrays are typically composed of $\sim~{\rm few}
\times 10^3$ dipoles (with approximately 10 per cent of the collecting
area and therefore sensitivity of the tentative SKA design) spread
over a region with maximum baseline $\sim 1$~km and designed to
operate at subsets of the frequency range $50 - 300$~MHz. Below we
focus on the possibility of detecting HII regions with LOFAR, but our
results are easily transferred to alternate designs for a
low-frequency interferometer.

\begin{figure}\label{snr_plot}
  \vspace{85mm} \includegraphics{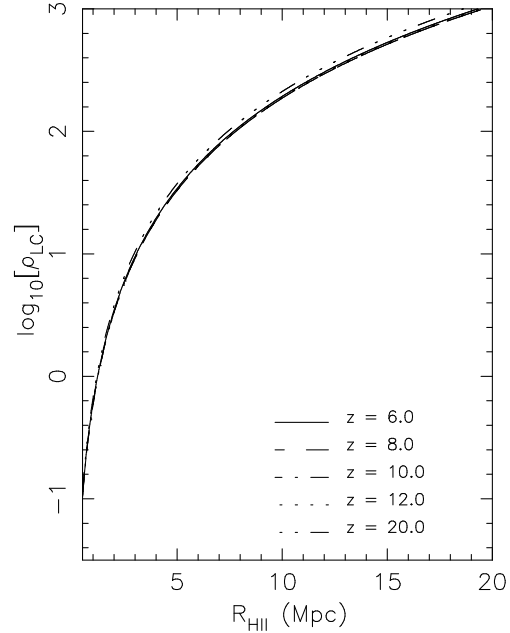}
  \caption{Signal to noise ratio as a function of physical radius for
    spherical HII regions at $z=6,8,10,12,20$. The calculation assumes
    constant sky noise at 250~K, 100 hours integration time, a
    spectral resolution of 0.1~MHz, a top hat beam with the same
    diameter as the source, and an observed bandwidth which
    encompasses the source.}
\end{figure}

\subsection{LOFAR sensitivity to HII regions}\label{LOFAR_sens}

\begin{figure*}\label{radius_dist}
  \vspace*{75mm} \includegraphics{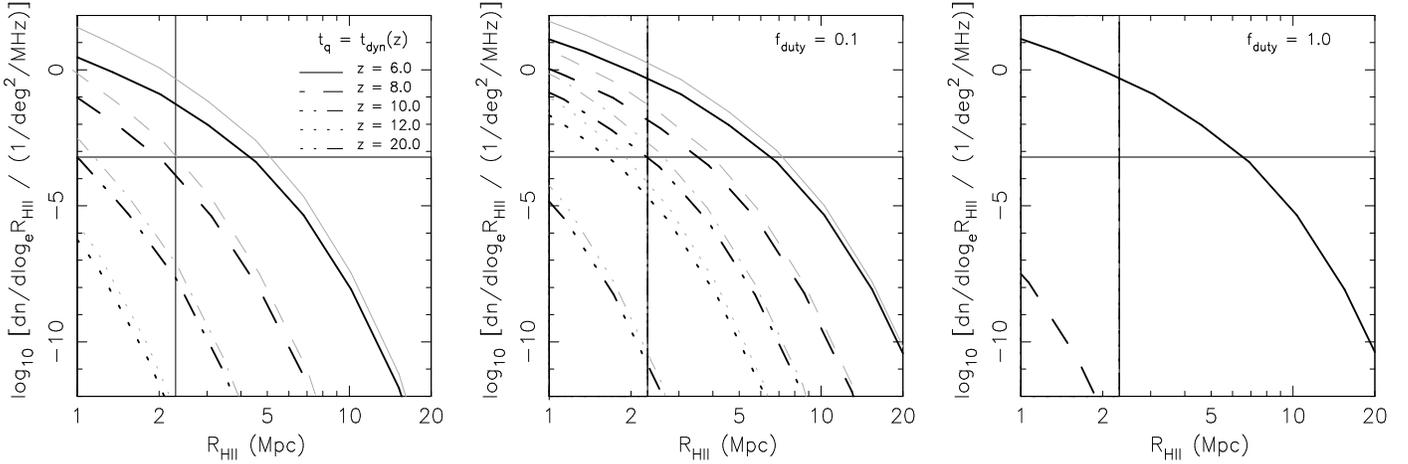}
\caption{The \emph{black} lines show the size distribution of 
  HII regions (physical radius) around active QSOs for $z =
  6,8,10,12,20$ for the QLFs in Figure 1. The \emph{grey} lines show
  the size distribution if fossil HII regions are included. The
  horizontal line corresponds to a source density of one per
  400~deg$^2$ per 4~MHz of observed bandwidth. The vertical line
  corresponds to the smallest HII region at $z=6$ detectable in 100
  hours of integration time, as calculated in
  \textsection\ref{LOFAR_sens}, however we find that the minimum
  detectable physical size changes very little with redshift.  Note
  that the surrounding IGM has been assumed to be neutral. }
\end{figure*}

We assume that the ionised region has a brightness temperature
contrast of $\delta T_{\rm HII} = 27$~mK~$[(1+z)/10]^{1/2}$,
corresponding to the limit where the hydrogen in the surrounding IGM 
is completely neutral, and the spin temperature ($T_s$) of the surrounding neutral gas is
efficiently coupled to the kinetic temperature of a warm IGM.

The compact LOFAR core, rather than the full array, will offer the
highest sensitivity to extended structures.  In the following we have
adopted array parameters corresponding to the published  design study
of LOFAR \footnote{http://www.lofar.org}. Assuming a spectral
resolution $\Delta \nu_{\rm ch}$ and integration time $\Delta \tau$,
an array of $N_{\rm stn}$ dual polarisation antennae, each with
collecting area $A_{\rm eff}$ and system temperature $T_{\rm sys}$,
will have a brightness temperature sensitivity  

\begin{eqnarray}\label{DelT}
\nonumber \Delta T_b (B_{\rm max})&=& \frac{\sigma_p B_{\rm max}^2}{2 k_{\rm
B}\Gamma}, \\ \nonumber&\approx& 128~{\rm mK} \left(\frac{T_{\rm sys}}{250~{\rm
K}}\right)\left( \frac{\Delta\nu_{\rm ch}}{0.1~{\rm
MHz}}\right)^{-1/2} \left( \frac{\Delta\tau}{100~{\rm
hr}}\right)^{-1/2} \\
&&\times \left[ \frac{B_{\rm max}}{2~{\rm km}}\right]^2
\left(\frac{N_{\rm stn}}{3200}\right)^{-1} \left[ \frac{A_{\rm
      eff}}{18~{\rm m}^2}\right]^{-1}\Gamma^{-1},
\end{eqnarray}

\noindent where $B_{\rm max}$ is the maximum distance between antennae in the 
core of the array, $\Gamma = \Omega_{\rm beam}/(\lambda/B_{\rm max})^2$ and
$\Omega_{\rm beam}$ is the solid angle of the synthesised beam. $k_B$
denotes the Boltzmann constant. The sensitivity to point sources $\sigma_p$
is given by,
\begin{eqnarray}\label{pt_source_rms}
\nonumber
\sigma_p&\approx&8.8\times10^{-5}~{\rm Jy}\left(\frac{T_{\rm
      sys}}{250~{\rm K}}\right)\left( \frac{\Delta\nu_{\rm
      ch}}{0.1~{\rm MHz}}\right)^{-1/2} \\ \nonumber
&&\times \left(\frac{\Delta\tau}{100~{\rm hr}}\right)^{-1/2}\left[ \frac{A_{\rm
      eff}}{18~{\rm m}^2}\right]^{-1}\left(\frac{N_{\rm
      stn}}{3200}\right)^{-1}.
\end{eqnarray}

\noindent 
LOFAR is expected to have a bandwidth of 4~MHz with a
spectral resolution of 4~kHz which will likely be binned into $\sim
0.1$~MHz channels. The spatial resolution of the LOFAR core will be
inversely proportional to the maximum distance between the antennae
($\theta_{\rm res} \approx \lambda/B_{\rm max} \sim 2.5~{\rm
  arcmins}\frac{(1+z)}{7}$). In comparison, the angular size
($\theta_{\rm HII}$) and frequency depth ($\Delta \nu_{\rm HII}$) of
an HII region with radius $R_{\rm HII}$ are given by

\begin{eqnarray}\label{ang_size}\nonumber
\theta_{\rm HII} &=& 2\frac{R_{\rm HII}}{D_c(z)}(1+z),\\
\backsimeq &14~{\rm arcmins}&\left( \frac{R_{\rm HII}}{5~{\rm Mpc}} \right) \left( \frac{1+z}{7} \right) \left[ \frac{D_c(z)}{D_c(z=6)} \right]^{-1},
\end{eqnarray}

\begin{eqnarray}\label{freq_size}\nonumber
\Delta \nu_{\rm HII} &=& 2\frac{H(z)}{c} \frac{R_{\rm HII}}{1+z} \nu_o, \\
\backsimeq &4.6~{\rm MHz}&\left( \frac{R_{\rm HII}}{5~{\rm
Mpc}}\right) \left( \frac{1+z}{7}\right)^{-1} \left[ \frac{H(z)}{H(z=6)}\right], 
\end{eqnarray}

\noindent where $\nu_o \approx 1420$~MHz is the rest-frequency of the 
signal and $D_c(z)$ is the comoving distance to a source at redshift
$z$. 

We now calculate the signal to noise ratio (SNR) with which a
spherical bubble of a given radius at a certain redshift can be
detected with $N_{\rm stn} = 3200$ antennae each with $A_{\rm eff} =
18$~m$^2$ in a circle of variable diameter $B_{\rm max}$.  The
brightness temperature sensitivity increases with the compactness of
the array.  The surface brightness sensitivity will be maximal when
the size of the beam is the same as the angular size of the source
$B_{\rm max} \sim \lambda_{\rm 21cm} / \theta_{\rm HII}$, $\Gamma
=1$. 

The SNR for a spherical HII region is then given by,

\begin{equation}
\rho_{\rm LC} \approx  \frac{\delta T_{\rm HII}} {\Delta T_b(B_{\rm max}
= \lambda_{\rm 21cm} / \theta_{\rm HII})}\left(\frac{\Delta \nu_{\rm  HII}}
{\Delta \nu_{\rm ch}} \right)^{1/2}\mathcal{F_{\rm HII}}.
\end{equation}

\noindent The factor $\mathcal{F}_{\rm HII}$ accounts for the fraction 
of the beam that is filled with ionised gas. For a spherical top heat
beam $\mathcal{F}_{\rm HII} = \frac{4/3 \pi R_{\rm HII}^3}{2 \pi
  R_{\rm HII}^3} = 2/3$. $\mathcal{F}_{\rm HII}$ may be adjusted for
different array patterns/synthesised beams and different HII region
geometries.

This prescription allows us to estimate the signal to noise ratio for
detection of an HII region as a function of its redshift and radius.
Figure~4 shows the calculated signal to noise ratio as a function of
the physical radius of an HII region at various redshifts.  Note that
the increase of SNR with size of the HII region will become shallower
once the size exceeds the beam size for the most compact configuration
possible.  This calculation indicates that the smallest bubble
surrounded by a neutral IGM that may be detected at the $5-\sigma$
level is $\sim 2.3$~Mpc at $z=6$. The minimum radius changes very
little with source redshift, and turns over at $z \sim 10$ due to the
dependence on the angular diameter distance.

Unfortunately, instrumental noise is not the only type of noise
relevant for the detection of HII regions. Residuals from the
subtraction of extragalactic and galactic foregrounds and the
correction of the distortion of the signal by the ionosphere will all
contribute to the noise. We currently have little idea how large the
noise due to these residuals will be.  The intensity fluctuations in
the 21cm emission due to density fluctuations of the IGM will also act
as noise (e.g. Kohler et al. 2005, Mellema et al. 2006). For
$100$~hour integrations with the LOFAR core the fluctuations due to
the IGM surrounding the source limit the sensitivity on scales larger
than $\sim 6$~arcminutes (Wyithe, Loeb \& Barnes~2005); increasing the
exposure time will not significantly decrease the minimum size of
detectable HII regions.  Furthermore, the edge of an HII region
surrounding a QSO is not expected to be sharp (see Zaroubi \&
Silk~2005) and the intrinsic shape will appear distorted due to light
travel time effects (see Wyithe \& Loeb~2004b; Yu~2005).  The signal
to noise ratios estimated here should thus be considered as (probably
rather optimistic) upper limits.

\subsection{The space density of HII regions}

In Figure 5 we show our predictions for the space density of HII
regions as a function of size at $z>6$ in units of ${\rm deg}^{-2}
{\rm MHz}^{-1}$ for each of the three models. The vertical and
horizontal lines show the typical size of a LOFAR observing volume
($\sim 400$~deg$^2$ and $4$~MHz of bandwidth) and our estimate for the
$5-\sigma$ detection threshold for the size. The effective
recombination timescale for hydrogen gas at $10^4$~K is comparable to
or somewhat shorter than the Hubble time.  As discussed by Wyithe,
Loeb \& Barnes~(2005) HII regions are thus expected to remain as
'fossils' for some time after the QSO turns off. We assume that the
HII regions continue to expand with the Hubble flow once the quasar
turns off and there are no further recombinations, and count the
number of fossil regions with physical radius R at $z$ by summing over
the size distribution function at redshift intervals corresponding to
the quasar lifetime. We assume that the quasars at $z=6$ are on
average half way through their optically luminous lifetime.  There are
essentially no fossil HII regions in the case of $f_{\rm duty} = 1.0$
since the QSO luminosity grows exponentially on a timescale comparable
to the quasar lifetime.  Note that for the other two models $1/f_{\rm
  fossil}(R,z)$ is $\sim 4$ times smaller than $1/f_{\rm lum}$ for the
smallest HII regions (corresponding to the faintest quasars, $L_{\rm
  bol} \sim 10^{11}~{\rm L}_{\sun}$) at $z = 6$ due to the rapid
evolution of the space density of QSOs.  $1/f_{\rm fossil}(R,z)$
asymptotes to one for large HII regions and high redshifts as
expected.  We show both the number of HII regions around active QSOs
(black lines) and the total number (including fossil bubbles) of HII
regions (grey lines).  We have assumed that the IGM is still neutral
all the way down to $z=6$.  We will come back to this point below.

For the merger-driven model we predict $\sim 14$ detectable sources
around active QSOs per LOFAR field at $z = 6$.  This number increases
to $\sim 340$ if we include fossil HII regions around
supermassive black holes which are no longer active.  The numbers are
similar to those reported by Wyithe, Loeb \& Barnes~(2005) for their
similar model.  In our merger driven model at least one detectable HII
region per LOFAR field is predicted out to $z \sim 7$.  Due to the
steep slope of the QLF at high luminosities, the numbers would
decrease more rapidly with redshift if the effective minimum
detectable radius is larger than we have estimated.

For our two passive evolution models the predictions are quite
different.  The model with the lower duty cycle offers a more
optimistic picture than the merger driven model.  The longer QSO
lifetime results in larger HII regions for the same QSO luminosity.
We predict $ \sim 135$ detectable HII regions per LOFAR field at $z
\sim 6$ around active luminous QSOs and detectable HII regions are
present up to $z \sim 10$.  The value of the maximum redshift depends
sensitively on the duty cycle and could be -- at least in principle --
much larger for smaller values of the duty cycle.  This would,
however, require the formation of very massive black holes at very
high redshift which would be difficult in most formation scenarios.

The passive evolution model with the higher value of the duty cycle
depicts a much more pessimistic scenario. The predicted numbers at
$z=6$ are similar to the other models, but the the presence of
detectable HII regions does not extend much beyond $z=6$. This is
easily explained by the very recent growth of the $z=6$ supermassive
black holes in this model.

Figure 6 shows the redshift evolution of detectable active and fossil
HII regions with $R_{\rm min} = 2.3$~Mpc.  Clearly the prospects for
detecting HII regions depend sensitively on how BHs grow at high
redshift.

So far we have assumed that the IGM is still neutral all the way down
to $z=6$.  However the neutral fraction at redshift $6$ is uncertain
and the Universe may already be highly ionised at $z\sim 6-7$
(Malhotra \& Rhoads~2005, Fan et al. 2006).  A smaller but still
significant neutral fraction would result in a larger HII region [see
equation~(\ref{lum-rad})], however the signal to noise ratio for the
detection of a resolved HII regions scales with the square root of the
neutral fraction ($ \rho_{\rm LC} \propto x_{\rm HI}^{1/2}$); sources
in a partially neutral IGM are more difficult to detect. 

For consistency we have calculated the volume filling factor of
hydrogen ionised by the QSOs in each of our models. We assume that our
model extends to QSOs as faint as $L_{\rm bol} = 10^{11}$~L$_{\sun}$,
include the contribution due to fossil regions and neglect
recombinations. Our filling factor should therefore represent an
upper limit on the expected value. We find a filling factor of HII
regions generated by the UV emission of active quasars in our models
which is less than $\sim 35$ per cent at $z \sim 6$ for the model with
the low duty fraction and for the merger model and $\sim 5$ cent for
the model with the large duty fraction.  For each model the filling
factor drops rapidly at higher redshift.  Note again that we have
integrated down to rather faint quasars and that taking recombinations
into account would also lower these values.  Our assumption that the
HII regions generated by QSOs are isolated may therefore be somewhat
questionable at $z \sim 6$ but should be reasonable at $z \ga 7$. Not
surprisingly the quasars in our models which are consistent with the
rather low observed space density of QSOs at $z\sim 6$ alone are
barely able to fully reionise hydrogen by $z\sim 6$.

The quasars should also contribute to the ionisation of hydrogen with
their X-ray emission. Due to their long mean free path, X-rays will
lead to a more uniform ionisation which includes the low-density IGM
far away from the quasars (e.g. Madau et al.~2004, Ricotti \&
Ostriker~2004).  The efficiency with which an X-ray photon ionises
hydrogen atoms in the IGM depends on the average number of ionisations
induced by secondary electrons, which is a function of the initial
photon energy as well as the existing ionised gas fraction (see Shull
\& van~Steenberg~1985).  We follow Ricotti \& Ostriker and assume that
each photon with energy $\gtrsim 1$~keV results in 10 ionisations in a
partially ionised IGM.  Assuming again that our model extends to
quasars as faint as $L_{\rm bol} = 10^{11}$~L$_{\sun}$ the mean
ionised fraction at $z=6$ due to the X-ray emission of active QSOs in
our model is less than 1 percent for all models.  

Another possibly relevant effect is the clustering of QSOs. Bright
sources are sufficiently rare that clustering should not be important.
For fainter sources overlap of clustered HII region could increase the
sizes compared to our estimates in particular if fossil HII regions
are taken into account.

\begin{figure}\label{radius_evo}
  \vspace{85mm} \includegraphics{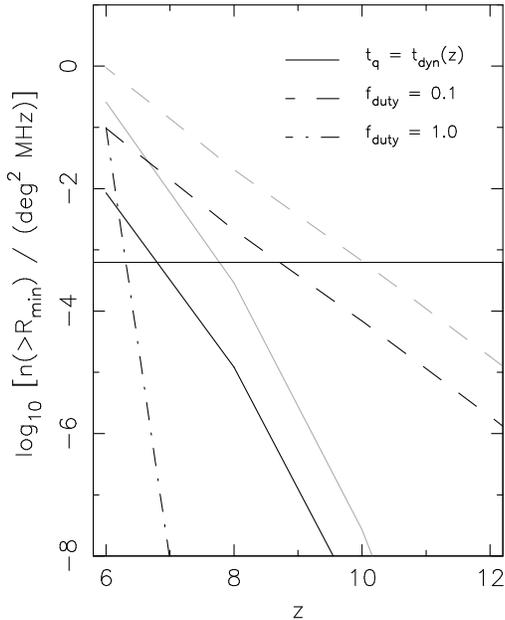}
  \caption{Number of HII regions larger than the calculated optimal detection limit of $R_{\rm min} = 2.3$~Mpc detectable limit as a function of redshift for the merger driven model (\emph{solid} line) and passive evolution models with $f_{\rm duty} = 0.1$ (\emph{dashed} line) and $f_{\rm duty} = 1$ (\emph{dot-dashed} line). The \emph{black} lines correspond to HII regions around active QSOs and the \emph{grey} lines include fossil HII regions. The horizontal line corresponds to one source per 400~deg$^2$ field of view per 4~MHz bandpass. Note that the surrounding IGM has been assumed to be neutral.}
\end{figure}

\section{Conclusions}

We have investigated the prospects of using the 21cm signature of the
regions around supermassive black holes at $z>6$ to probe the growth
history of these supermassive black holes and the ionisation history
of their surrounding IGM.

We have presented three simple, physically motivated models for the
high redshift growth history of supermassive black holes 
in which the growth is either  simple Eddington limited accretion 
or due to short-lived accretion phases triggered by the merger of
galaxies.  We have used a simple analytical model for the size of HII 
regions in order to generate predictions for the size distribution of 
HII regions for each of our model QLFs. Our results show that  
models which are all consistent with the luminosity function of 
bright QSOs at $z = 6$ but differ in the growth history of the 
supermassive black holes give very different results for the predicted 
number of detectable HII  regions.

Our estimates for the number of detectable HII regions at $z \sim 6$
are promising, up to several hundred detectable sources per LOFAR
field around active QSOs and possibly many more fossil regions.
However, the number of detectable HII regions depends sensitively on
unknown details of SMBH growth and the ionisation state of the IGM. If
the QLF evolves to $z >6$ as predicted in our merger-driven model,
then isolated HII regions will be prominent features of the high
redshift neutral hydrogen distribution providing reionisation was
completed late ($z < 7$).  However if the SMBH population is growing
rapidly due to continuous accretion, or reionisation occurred much
earlier than $z \sim 6$, then the detection of QSO HII regions may
prove elusive.

The most optimistic picture emerges if supermassive black holes grow
more slowly over an extended period of time as in our Eddington
limited accretion model with a smaller duty cycle ($f_{\rm duty}=
0.1$). In this case detectable sources may be present to very large
redshifts and the prospects for detection are good even if
reionisation is completed relatively early.

Observation of future 21cm experiments like LOFAR should  not only
provide a wealth of information on how the Universe was reionised but
may also provide important clues for the growth history of
supermassive black holes at very high redshift.

\section*{Acknowledgements}

The authors wish to thank Stuart Wyithe for useful comments on a draft
and the referee for suggesting calculating the HII region filling
factor and the degree of pre-reionisation by X-rays.  KJR is supported
by an Overseas Research Scholarship and the Cambridge Australia Trust.
This research was supported in part by the National Science
Foundation under Grant No. PHY99-07949.

\bsp

\label{lastpage}

\end{document}